\RequirePackage{fix-cm}
\documentclass[twocolumn]{svjour3}          
\smartqed  
\pdfoutput=1
\usepackage{graphicx}
\usepackage[colorlinks,plainpages=false,linkcolor=black,urlcolor=blue,citecolor=blue,pdfpagemode=UseNone,pdfstartview=FitH]{hyperref}
\begin{document}

\title{Electronic band structure of ferro-pnictide superconductors from ARPES experiment}

\titlerunning{Electronic structure of ferro-pnictides from ARPES}        

\author{A.~A.~Kordyuk \and
        V.~B.~Zabolotnyy \and
        D.~V.~Evtushinsky \and
        A.~N.~Yaresko \and
        B.~B\"{u}chner \and
        S.~V.~Borisenko
}


\institute{A.~A.~Kordyuk \at
           Institute of Metal Physics of National Academy of Sciences of Ukraine, 03142 Kyiv, Ukraine \\
           \email{kordyuk@gmail.com}           
           \and
           A.~A.~Kordyuk \and V.~B.~Zabolotnyy \and D.~V.~Evtushinsky \and B.~B\"{u}chner \and S.~V.~Borisenko \at
           Institute for Solid State Research, IFW-Dresden, P.O.Box 270116, D-01171 Dresden, Germany
           \and
           A.~N.~Yaresko \at
           Max-Planck-Institut f\"{u}r Festk\"{o}rperforschung, Heisenbergstra{\ss}e 1, 70569 Stuttgart, Germany
}

\date{Received: date / Accepted: date}

\maketitle

\begin{abstract}
ARPES experiments on iron based superconductors show that the differences between the measured and calculated electronic band structures look insignificant but can be crucial for understanding of the mechanism of high temperature superconductivity. Here we focus on those differences for 111 and 122 compounds and discuss the observed correlation of the experimental band structure with the superconductivity.

\end{abstract}

\section{Introduction}
\label{intro}

\begin{figure}[t]
\begin{center}
\includegraphics[width=.49\textwidth]{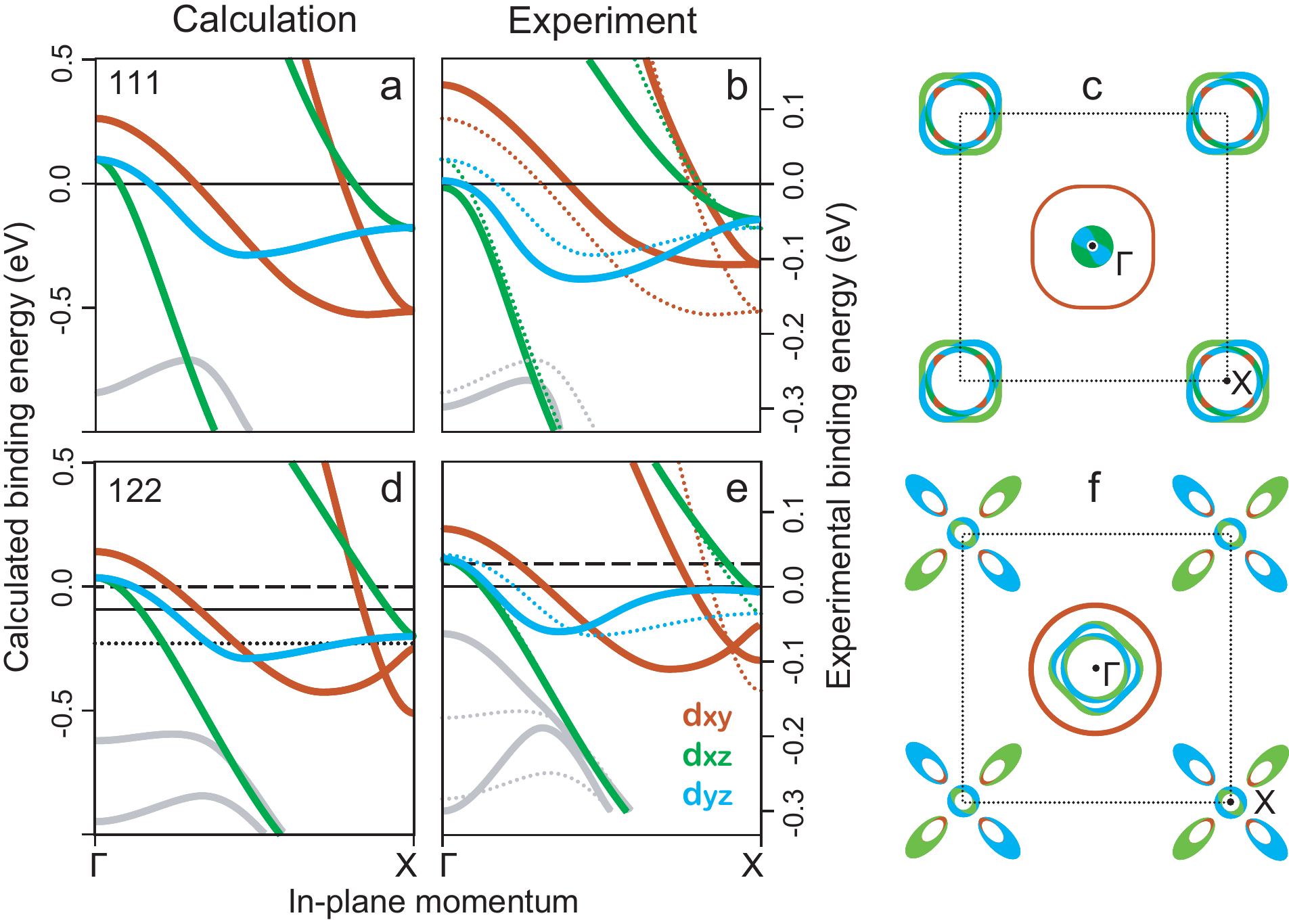}
\caption{\label{Fig_bands} Electronic band structure of LiFeAs (a-c), a representative 111 compound, and BaFe$_2$As$_2$ (BFA) / Ba$_{0.6}$K$_{0.4}$Fe$_2$As$_2$ (BKFA) (d-f), the parent/optimally doped 122 compound: the electronic bands, calculated (a, d) and derived from ARPES experiment (b, e), and the Fermi surfaces of LiFeAs (c) and BKFA (f), as seen by ARPES. The bands and FS contours are colored by the most pronounced orbital character: Fe 3$d_{xy}$, 3$d_{xz}$, and 3$d_{yz}$.}
\end{center}
\end{figure}

One can safely say that the visiting card of the iron based superconductors is their complex electronic band structure that usually results in five Fermi surface sheets (see Fig.\,\ref{Fig_bands}): three around the center of the Fe$_2$As$_2$ Brillouin zone and two around the corners. Band structure calculations predict rather similar electronic structure for all the ferro-pnictides and ferro-chalcogenides (see \cite{AndersenAdP2011,Sadovskii} and references therein) and the angle resolved photoemission spectroscopy (ARPES) \cite{DamascelliRMP2003}, the most direct tool to see the real electronic band structure of crystals, shows that it is indeed the case: one can fit the calculated bands to the experiment if it is allowed to renormalize them about 3 times and shift slightly with respect to each other \cite{YiPRB2009,DingJoPCM2011,BorisenkoPRL2010,BorisenkoJPCS2011}.

As a consequence of such a complex band structure, in which several van Hove singularities (vHs) stay close to the Fermi level, the electronic properties of iron based superconductors, as a function of doping, pressure, and the temperature, should be swarm with crossovers. Therefore, it is tempting to build a general phase diagram of these compounds based on their common band structure and find the correlations of this structure with superconductivity. In this paper, considering the most ``ARPESable" 111 and 122 compounds, we summarize the differences of their experimental and calculated band structures, show that these differences can be important for understanding of the pairing mechanism, and, building the generalized phase diagram, discuss the observed correlations of the experimental band structure with superconductivity.


\section{LiFeAs: no nesting but Lifshitz transition}
\label{111}


Among the iron based superconductors the most ``arpesable'' compound is LiFeAs \cite{BorisenkoPRL2010}. It cleaves between the two Li layers, thus revealing a non-polar surface with protected topmost FeAs layer; it is stoichiometric, i.e. impurity clean; it has the transition temperature about 18\,K and one can measure the superconducting gap by ARPES and compare its value to bulk techniques; it is non-magnetic and, consequently, the observed band structure is free of SDW replicas; and, finally, its electronic bands are the most separated from each other that allows one to disentangle them most easily and analyse their fine structure \cite{KordyukPRB2011}.

In Fig.\,\ref{Fig_bands}(a) we show a fragment of the low energy electronic band structure of LiFeAs calculated using the LMTO method in the atomic sphere approximation \cite{AndersenPRB1975}. The same calculated bands but 3 times renormalized are repeated in panel (b) by the dotted lines to compare with the dispersions derived from the numerous ARPES spectra \cite{BorisenkoPRL2010,KordyukPRB2011} shown in the same panel by the thick solid lines \cite{SI}. The experimental Fermi surface is sketched in panel (c). The five bands of interest are colored in accordance to the most pronounced orbital character: Fe 3$d_{xy}$, 3$d_{xz}$, and 3$d_{yz}$ \cite{LeePRB2008,GraserNJoP2009}. Those characters have helped us to identify uniquely the bands in the experimental spectra using differently polarized photons \cite{BorisenkoPRL2010}.

Comparing the results of the experiment and renormalized calculations, one can see that the strongest difference is observed around $\mathrm{\Gamma}$ point: the experimental $d_{xy}$ band is shifted up about 40 meV (120 meV, in terms of the bare band structure) while the $d_{xz}$/$d_{yz}$ bands are shifted about 40 (120) meV downwards. Around the corners of the BZ (X point) the changes are different, the up-shift of the $d_{xy}$ band in X point is about 60 meV while the $d_{xz}$/$d_{yz}$ bands are also shifted up slightly (about 10 meV). At the Fermi level, the largest hole-like FS sheet around $\mathrm{\Gamma}$ point, formed by $d_{xy}$ band, is essentially larger in experiment than in calculations. This is compensated by the shrunk $d_{xz}$/$d_{yz}$ FSs where the larger one has become three-dimensional, i.e. closed also in $k_z$ direction, and the smallest one has disappeared completely. The electron-like FSs have changed only slightly, alternating its character in $\mathrm{\Gamma}$X direction due to shift of the crossing of $d_{xz}$ and $d_{xy}$ bands below the Fermi level, see Fig.\,\ref{Fig_bands}(b). So, the experimental electronic band structure of LiFeAs has the following very important differences from the calculated one  \cite{BorisenkoPRL2010}: (i) there is no FS nesting, and (ii) the vHs, the tops of the $d_{xz}$/$d_{yz}$ bands at $\mathrm{\Gamma}$ point, stays in the vicinity of the Fermi level, i.e. the system is very close to a Lifshitz transition \cite{LifshitzZETF1960}. The latter makes the band structure of LiFeAs similar to the structure of optimally doped Ba(Fe$_{1-x}$Co$_{x})$As$_2$ (BFCA) \cite{LiuPRB2011}, as discussed below.

\section{Propeller-like structure in 122}
\label{122}

\begin{figure}[t]
\begin{center}
\includegraphics[width=.49\textwidth]{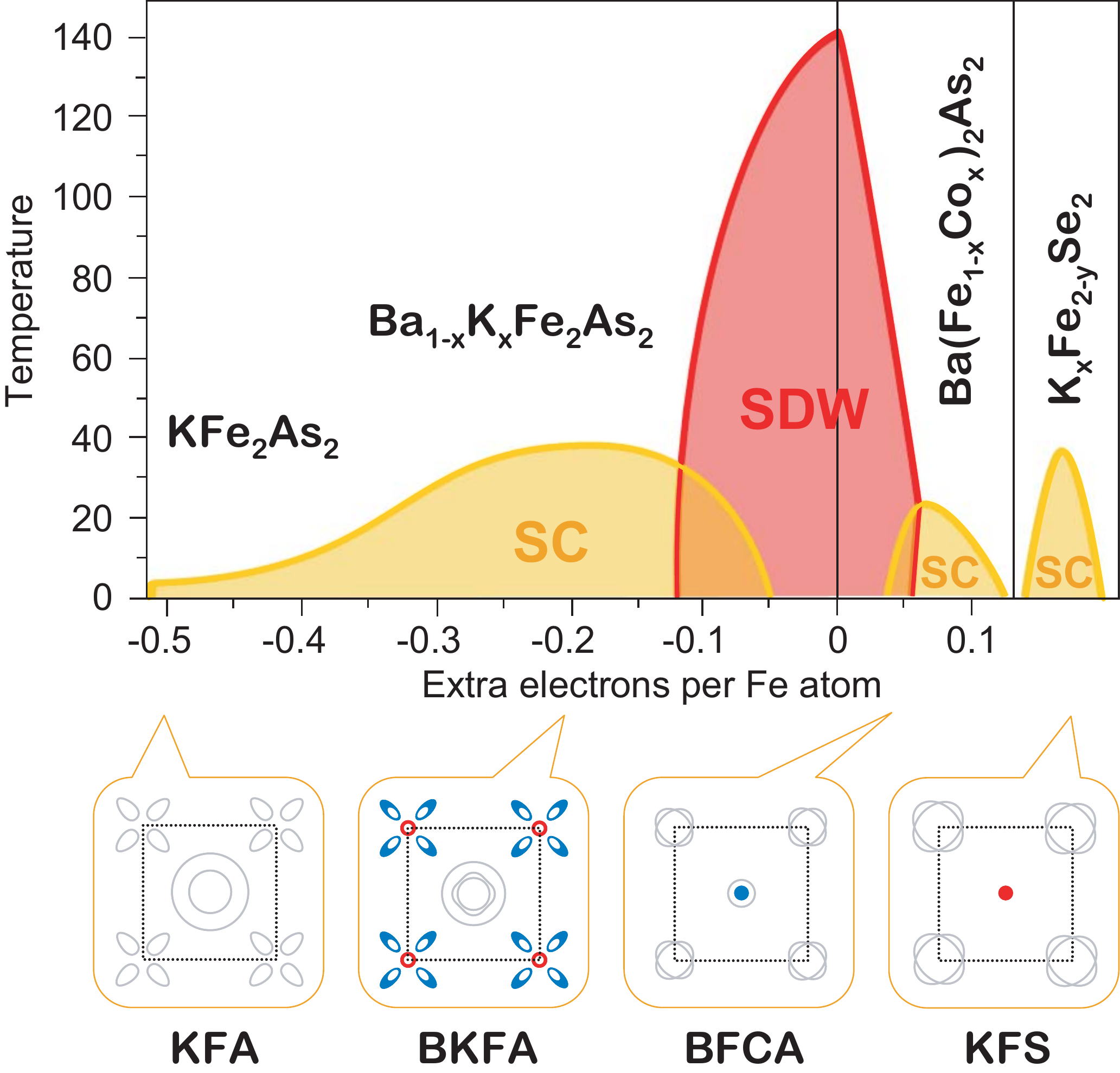}
\caption{\label{PhD}{PhD}. Phase diagram of the 122 family of ferro-pnictides complemented by the 122(Se) family as a generalized band structure driven diagram for the iron based superconductors. The insets show that the Fermi surfaces for every compound close to $T_{c\mathrm{max}}$ are in the proximity of Lifshitz topological transitions: the corresponding FS sheets are highlighted by color (blue for hole- and red for electron-like).}
\end{center}
\end{figure}

\begin{figure}[t]
\begin{center}
\includegraphics[width=.49\textwidth]{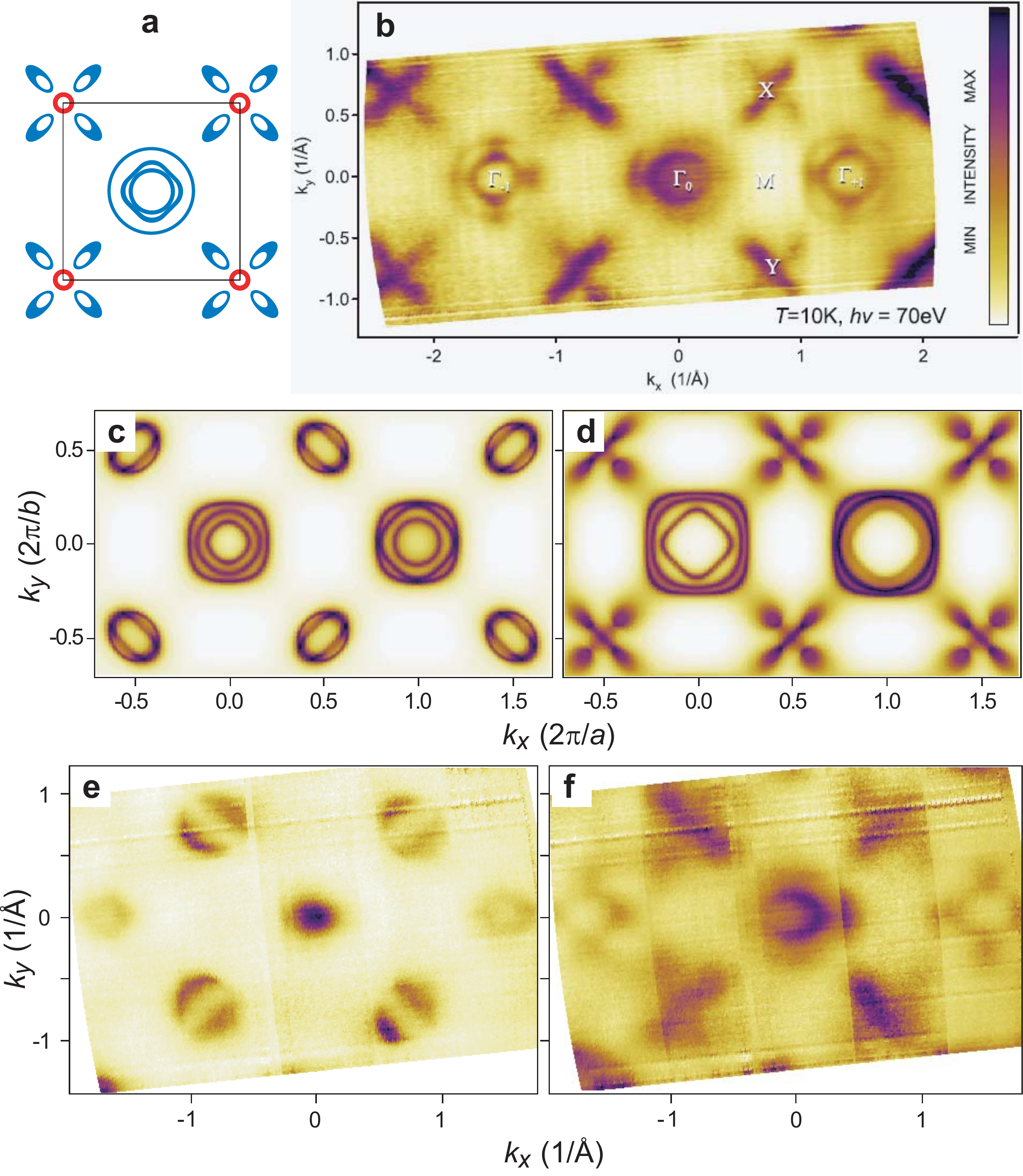}
\caption{\label{Fig_PropoMaps} Propeller-like Fermi surface in 122. The sketch (a) of the FS of the optimally doped BKFA is based on the ARPES map (b). The cuts of the calculated and renormalized band structure of BFA at $-30$ meV (c) and $-76$ meV (d). The cuts of ARPES spectrum of BFCA at $E_F$ and $-90$ meV.}
\end{center}
\end{figure}

Due to one alkaline earth metal atom per formula unit, the 122 family of ferro-pnictides does not have such an easy termination plane as the 111 family does and, therefore, might not be so perfect for ARPES. Nevertheless, the 122 family is the most studied by ARPES. The main reason for this is the variety of high quality crystals of different compounds with wide ranges of doping in both hole and electron sides \cite{WenARoCMP2011} that form a rich phase diagram (see Fig.\,\ref{PhD}) where the superconductivity and magnetism compete or coexist. In addition, it has appeared that the ARPES spectra well represent the bulk electronic structure of this family, at least, for the hole doped Ba$_{1-x}$K$_{x}$Fe$_2$As$_2$ (BKFA) and Ba$_{1-x}$Na$_{x}$Fe$_2$As$_2$ (BNFA), where the superconducting gap is routinely observed \cite{DingEPL2008,EvtushinskyPRB2009,Evtushinsky2011} and is in a good agreement with the bulk probes \cite{EvtushinskyNJP2009}. This poses the 122 family as the main arena to study the rich physics of the iron-based superconductors.

Here we focus on the parent stoichiometric BaFe$_2$As$_2$ (BFA), the electron doped Ba(Fe$_{1-x}$Co$_{x})$As$_2$ (BFCA), and the hole doped BKFA or BNFA. Another stoichiometric compound, KFe$_2$As$_2$ (KFA), is considered as an extremely overdoped one with 0.5 holes per Fe atom. A representative fragment of the calculated electronic band structure of BFA is shown in Fig.\,\ref{Fig_bands}(d). It is very similar to the band structure of LiFeAs with a small complication at the bottom of the $d_{xy}$ bands in X point that is a consequence of body-centered tetragonal stacking of FeAs layers instead of simple tetragonal stacking in LiFeAs.

With the highest, in 122 family, transition temperature ($T_c$ = 38~K) and the sharpest ARPES spectra, the hole doped BKFA and BNFA are the most promising and the most popular objects for trying to understand the mechanism of superconductivity in ferro-pnictides. This said, it is important to stress that the FS of the optimally doped Ba$_{0.6}$K$_{0.4}$Fe$_2$As$_2$ and Ba$_{0.6}$Na$_{0.4}$Fe$_2$As$_2$ is topologically different from the expected one: instead of two electron-like pockets around the corners of the Fe$_2$As$_2$ BZ (X and Y points) there is a propeller-like FS with the hole-like blades and a very small electron-like center \cite{ZabolotnyyN2009,ZabolotnyyPhC2009}, as shown in Fig.\,\ref{Fig_PropoMaps}(a,b). Curiously enough, despite the despite the experimental reports of the propeller like FS, the ``parent'' FS is still used in a number of theoretical models and as a basis for interpretation of experimental results such as superconducting gap symmetry.

Our first interpretation of the propeller-like FS, as an evidence for an additional electronic ordering \cite{ZabolotnyyN2009}, was based on temperature dependence of the photoemission intensity around X point and on the similarity of its distribution to the parent BFA, but the interpretation based on a shift of the electronic band structure \cite{YiPRB2009} was also discussed. Now, while it seems that the electronic ordering plays a certain role in spectral weight redistribution \cite{EvtushinskyJPSJ2011}, we have much more evidence for the ``structural'' origin of the propellers: (1) The propeller-like FS, such as shown Fig.\,\ref{Fig_PropoMaps}(a), is routinely observed for every optimally doped BKFA or BNFA crystals we have studied. (2) In extremely overdoped KFA \cite{SatoPRL2009,YoshidaJPCS2011}, where the magnetic ordering is not expected at all, they naturally (according to rigid band approximation) evolve to larger hole-like propellers. (3) One can see the same propellers in the spectrum of the overdoped ($T_c =$ 10\,K) BFCA at 90 meV below the Fermi level (see Fig.\,\ref{Fig_PropoMaps}(f)).

In fact, one can get very similar distribution of the spectral weight observed by ARPES in a model based on LDA calculations. In Fig.\,\ref{Fig_PropoMaps}(c,d) we model the Fermi surface maps within the rigid band approximation starting from the calculated BFA band structure: the energy cuts of the 3D band structure are shown for $k_z = 0$ but integrated in the window $\pm0.5|\Gamma\mathrm{Z}|$. The shift of the chemical potential to $-90$ meV (or 30 meV by renormalized scale of binding energy), which is shown by the solid horizontal line in Fig.\,\ref{Fig_bands}(d) and corresponds to the optimally doped BKFA ($x = 0.4$ or 0.2 holes per Fe atom), gives the FS shown in Fig.\,\ref{Fig_PropoMaps}(c), which is topologically equivalent to the parent one. The larger shift down to $-228$ meV (renormalized 76 meV), shown by a dotted horizontal line in Fig.\,\ref{Fig_bands}(d), results in the topologically different FS as shown in the intensity map in Fig.\,\ref{Fig_PropoMaps}(d), that is very similar to the one observed by ARPES.

In Fig.\,\ref{Fig_bands}(e) we show the experimental bands (solid lines), derived from a number of ARPES spectra, on top of the bands (thin dotted lines) calculated for parent BFA, 3 times renormalized, and shifted by 30 meV, as discussed above, to model the band structure expected for Ba$_{0.6}$K$_{0.4}$Fe$_2$As$_2$. One can see that the difference between the experimental and ``expected'' dispersions is even smaller than in case of LiFeAs and mainly appears near X point as 40 meV shifts of the $d_{xz}$/$d_{yz}$ bands and one of $d_{xy}$ bands. These small shifts, however, result in the topological Lifshitz transition of the FS and the question is how it is related to superconductivity. This brings us to the last section of the paper.

\section{Phase diagram and band structure}
\label{S_PhD}

\begin{figure}[t]
\begin{center}
\includegraphics[width=.49\textwidth]{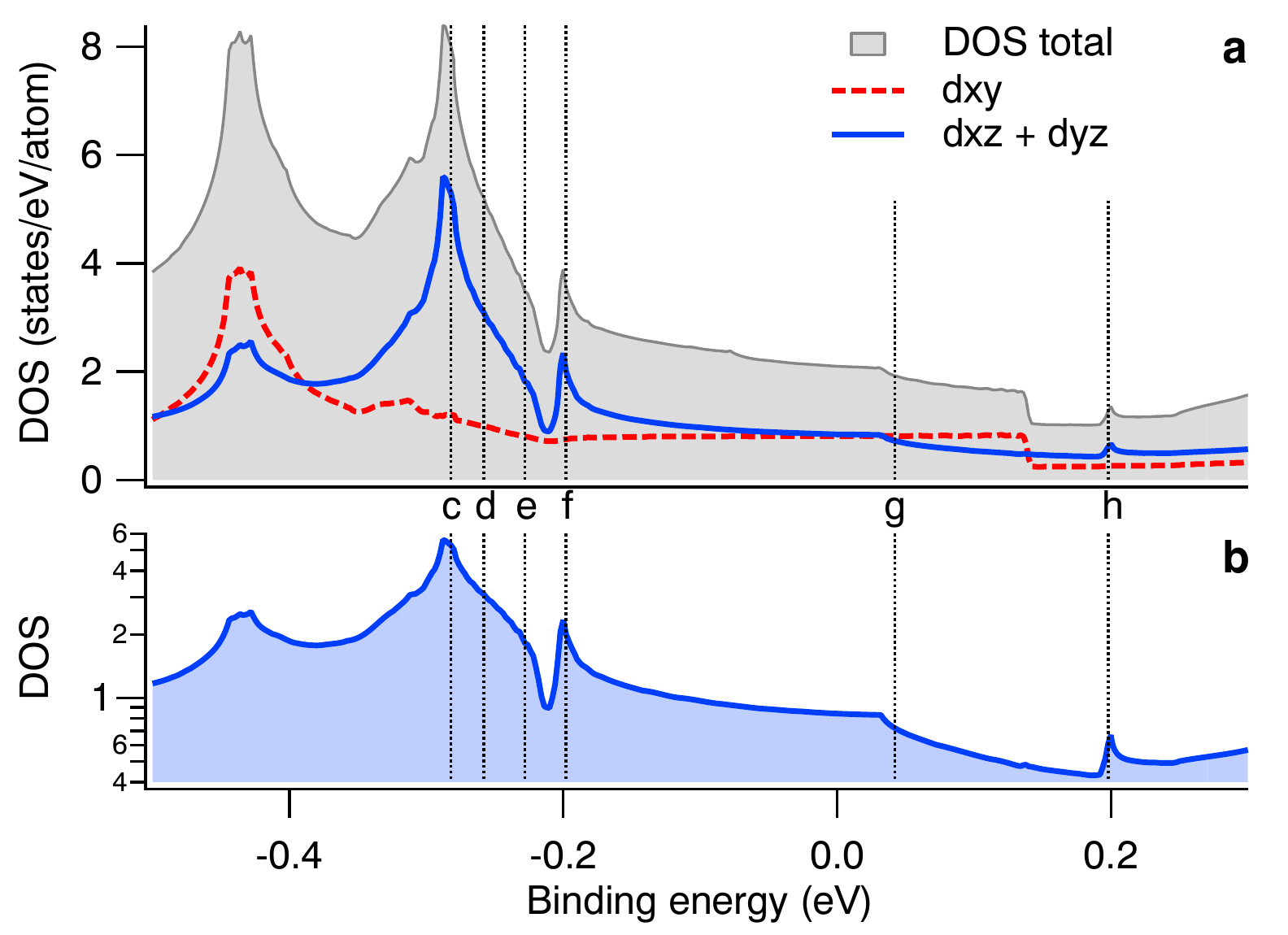}
\includegraphics[width=.49\textwidth]{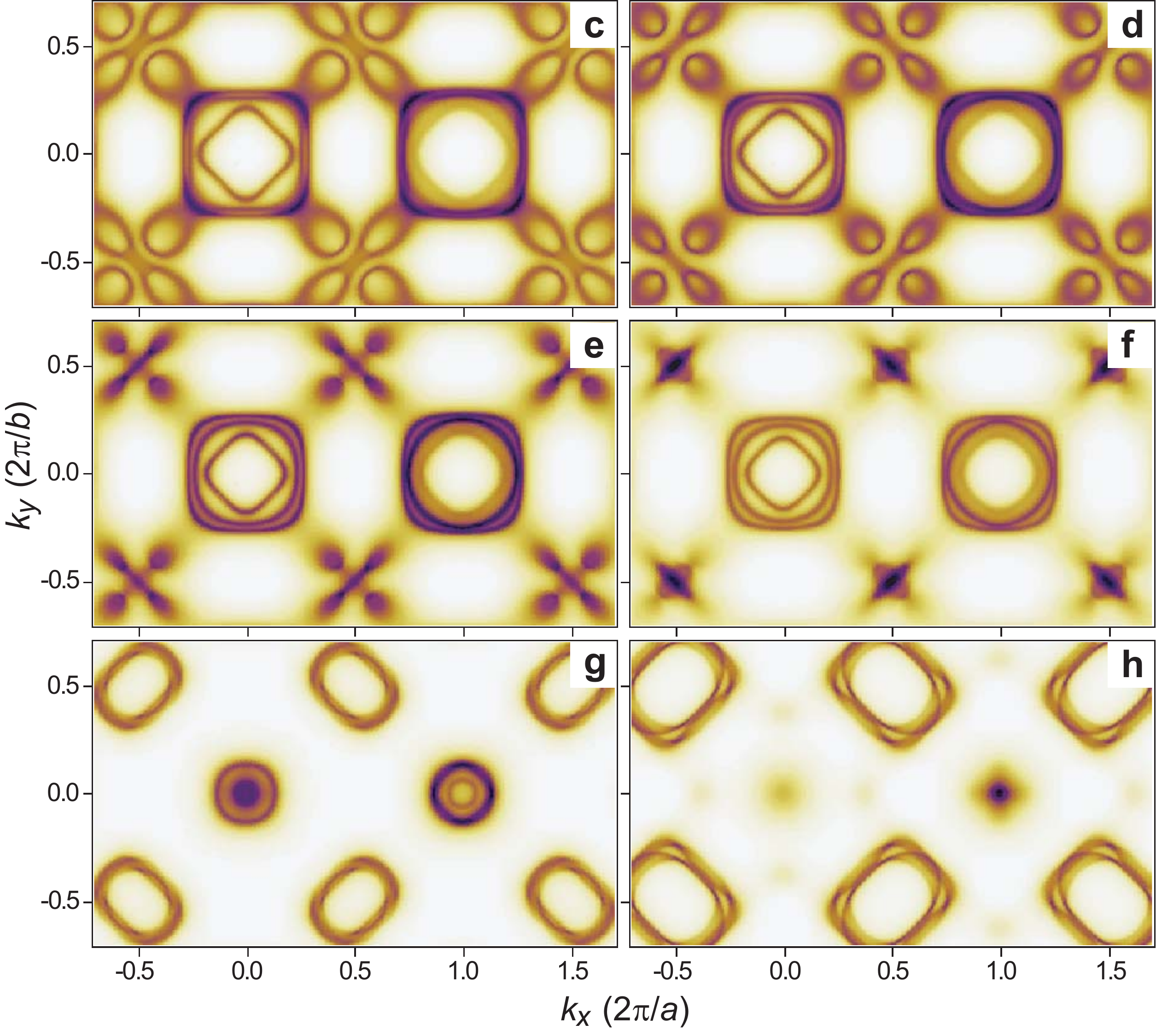}
\caption{\label{DOS} Electronic density of states (DOS) of parent BFA (a,b) (total and orbitally resolved) and the model Fermi surfaces (c-h) that would correspond to the positions of the chemical potential marked on panel (b) by vertical lines. }
\end{center}
\end{figure}

Naturally, one would like to examine whether the peak in the electronic density of states (DOS), related to the Lifshitz transition, can be responsible for the enhancement of superconductivity in BKFA. In Fig.\,\ref{DOS} we show DOS calculated for the parent BFA (a,b) and the model Fermi surfaces (c-h) that would correspond to the positions of the chemical potential marked on panels (a,b) by the vertical dotted lines. One can see that the chemical potential, for which the FS shown in panel (e) would be the most similar to the experimental FS of BKFA \cite{ZabolotnyyN2009,ZabolotnyyPhC2009}, drops in the region where DOS of $d_{xz}$/$d_{yz}$ bands exhibits singularities. Strictly  speaking, at the energy of $-228$ meV, to which panel (e) corresponds, DOS is not peaked but is increasing with lowering energy, hinting that a simple correlation between DOS and $T_c$, as suggested in \cite{Sadovskii}, does not work for BKFA. One can argue that the experimental $d_{yz}$ band is much flatter than the calculated one, see Fig.\,\ref{Fig_bands}(e), which should result in the enhancement of DOS at the Fermi level. Also, one can speculate that the normal state FS of optimally doped BKFA is more close to the case (f) from Fig.\,\ref{DOS} and transforms to (e) as a result of the electronic ordering. Nevertheless, accepting direct correlation between DOS and $T_c$, one would have a problem to explain why the extremely doped KFA, represented here by panel (d), has much higher DOS, but much lover $T_c = 3$\,K. On the other hand, the high $T_c$ superconductivity scenario driven by interband pairing in a multiband system in the proximity of a Lifshitz topological transition \cite{InnocentiPRB2010,InnocentiSUST2011}, looks more promising alternative for BKFA. This  said, it seems extremely challenging task for chemists to go with overdoping still further in order to reach the $d_{xz/yz}$ saddle points (c) responsible for the largest DOS peak at $-282$ eV. Interestingly, the same can be suggested for LiFeAs, where DOS \cite{SI} shows a much higher peak of the same $d_{xz/yz}$ origin as for panel (f) in BFA.

Going back to the Lifshitz transitions in iron based superconductors, let us overview their electronic band structures now accessible by ARPES. Recently, the correlation of the Lifshitz transition with the onset of superconductivity has been observed in BFCA \cite{LiuPRB2011,LiuNP2010}. The study has been mainly concentrated on the outer hole-like FS formed by $d_{xy}$ orbitals, nevertheless, it has been also found \cite{LiuPRB2011} that the tops of the $d_{xz}$/$d_{yz}$ bands go to the Fermi level for the samples with the optimal doping and $T_c = 24$\,K. Thus, the FS of optimally doped BFCA is similar to the one shown in Fig.\,\ref{DOS}(g) where the $\mathrm{\Gamma}$-centered $d_{xz/yz}$ FS sheet is in the proximity of a Lifshitz transition. If only the $d_{xz}$/$d_{yz}$ bands are concerned, the case of LiFeAs is very similar, as has been shown earlier \cite{BorisenkoPRL2010} and discussed above. One can add another 111 compound here, NaFeAs, that also has the tops of $d_{xz}$/$d_{yz}$ bands very close to the Fermi level \cite{HePRL2010}, though its electronic structure is complicated by the magnetic ordering.

One more example to support this picture comes from the iron selinides, which form an important family (known as 122(Se) or 245) of the iron based superconductors with purely electron-like FS and the highest transition temperature about 31\,K (see \cite{Sadovskii} and references therein). The ARPES spectra from these compounds \cite{WangEPL2011} are not very sharp yet, but one can confidently say that the bottom of the electron pocket at the center of the BZ is very close to the Fermi level, that allows us to associate this family with the FS shown in Fig.\,\ref{DOS}(h) and place them on the electron overdoped side of the generalized phase diagram, as shown in Fig.\,\ref{PhD}. At the end we note that in all known cases the bands those Lifshitz transitions do correlate with $T_c$ have dominantely Fe 3$d_{xz/yz}$ orbital character.

\section{Conclusions}
Considering all the electronic band structures of the iron based superconductors that can be derived from ARPES we have found that the Fermi surface of every optimally doped compound (the compounds with highest $T_c$) has the Van Hove singularities of the Fe 3$d_{xz/yz}$ bands in the vicinity to the Fermi level. This suggests that the proximity to an electronic topological transition, known as Lifshitz transition for one of the multiple Fermi surfaces, may be very important in these compounds for controlling interband pairing in multigap superconductivity, as it was recently suggested \cite{InnocentiPRB2010,InnocentiSUST2011}. Based on this empirical observation, we predict that hole overdoping of KFe$_2$As$_2$ and LiFeAs compounds is a possible way to increase the $T_c$.

\begin{acknowledgements}
We acknowledge discussions with A. Bianconi, A.V. Boris, A.V. Chubukov, A.M. Gabovich, D.S. Inosov, Yu.V. Kopaev,  M.M. Korshunov, I.V. Morozov,  I.A. Nekrasov, S.G. Ovchinnikov, M.V. Sadovskii, and M.A. Tanatar. The project was supported by the DFG priority program SPP1458, Grants No. KN393/4, BO1912/2-1.
\end{acknowledgements}



\end{document}